\documentclass[letter]{aa} 
\usepackage{graphicx,natbib}
\usepackage{txfonts}
\begin{document}
   \title{The radio delay of the exceptional \object{3C 454.3} outburst}

   \subtitle{Follow-up WEBT observations in 2005--2006\thanks{For questions
   regarding the availability of the data presented in this paper,
   please contact Massimo Villata ({\tt villata@oato.inaf.it}).}}

\author{M.~Villata \inst{1}
\and C.~M.~Raiteri \inst{1}
\and M.~F.~Aller \inst{2}
\and U.~Bach \inst{1,3}
\and M.~A.~Ibrahimov \inst{4}
\and Y.~Y.~Kovalev \inst{3,5,6}
\and O.~M.~Kurtanidze \inst{7,8,9}
\and V.~M.~Larionov \inst{10,11}
\and C.-U.~Lee \inst{12}
\and P.~Leto \inst{13}
\and A.~L\"ahteenm\"aki \inst{14}
\and K.~Nilsson \inst{15}
\and T.~Pursimo \inst{16}
\and J.~A.~Ros \inst{17}
\and N.~Sumitomo \inst{18}
\and A.~Volvach \inst{19}
\and H.~D.~Aller \inst{2}
\and A.~Arai \inst{18}
\and C.~S.~Buemi \inst{20}
\and J.~M.~Coloma \inst{17}
\and V.~T.~Doroshenko \inst{21}
\and Yu.~S.~Efimov \inst{22}
\and L.~Fuhrmann \inst{1,23,3}
\and V.~A.~Hagen-Thorn \inst{10,11}
\and M.~Kamada \inst{18}
\and M.~Katsuura \inst{18}
\and T.~Konstantinova \inst{10}
\and E.~Kopatskaya \inst{10}
\and D.~Kotaka \inst{18}
\and Yu.~A.~Kovalev \inst{5}
\and M.~Kurosaki \inst{18}
\and L.~Lanteri \inst{1}
\and L.~Larionova \inst{10}
\and M.~G.~Mingaliev \inst{24}
\and S.~Mizoguchi \inst{18}
\and K.~Nakamura \inst{18}
\and M.~G.~Nikolashvili \inst{7}
\and S.~Nishiyama \inst{18}
\and K.~Sadakane \inst{18}
\and S.~G.~Sergeev \inst{22}
\and L.~A.~Sigua \inst{7}
\and A.~Sillanp\"a\"a \inst{15}
\and R.~L.~Smart \inst{1}
\and L.~O.~Takalo \inst{15}
\and K.~Tanaka \inst{18}
\and M.~Tornikoski \inst{14}
\and C.~Trigilio \inst{20}
\and G.~Umana \inst{20}
}

   \offprints{M.\ Villata}

\institute{INAF, Osservatorio Astronomico di Torino, Italy\\
              \email{villata@oato.inaf.it}
\and Department of Astronomy, University of Michigan, MI, USA
\and Max-Planck-Institut f\"ur Radioastronomie, Germany
\and Ulugh Beg Astronomical Institute, Academy of Sciences of Uzbekistan, Uzbekistan
\and Astro Space Center of Lebedev Physical Institute, Russia
\and National Radio Astronomy Observatory, Green Bank, WV, USA
\and Abastumani Astrophysical Observatory, Georgia
\and Astrophysikalisches Institut Potsdam, Germany
\and Landessternwarte Heidelberg-K\"onigstuhl, Germany
\and Astronomical Institute, St.-Petersburg State University, Russia
\and Isaac Newton Institute of Chile, St.-Petersburg Branch
\and Korea Astronomy and Space Science Institute, South Korea
\and INAF, Istituto di Radioastronomia Sezione di Noto, Italy
\and Mets\"ahovi Radio Observatory, Helsinki University of Technology, Finland
\and Tuorla Observatory, Finland
\and Nordic Optical Telescope, Roque de los Muchachos Astronomical Observatory, TF, Spain
\and Agrupaci\'o Astron\`omica de Sabadell, Spain
\and Astronomical Institute, Osaka Kyoiku University, Japan
\and Radio Astronomy Laboratory of Crimean Astrophysical Observatory, Ukraine
\and INAF, Osservatorio Astrofisico di Catania, Italy
\and Moscow State University, Russia
\and Crimean Astrophysical Observatory, Ukraine
\and Dipartimento di Fisica e Osservatorio Astronomico, Universit\`a di Perugia, Italy
\and Special Astrophysical Observatory, Russia
}

   \date{   }

 
  \abstract
   {In spring 2005 the blazar \object{3C 454.3} was observed in an unprecedented bright state
   from the near-IR to the hard X-ray frequencies. A mm outburst peaked in June--July 2005, and
   it was followed by a flux increase at high radio frequencies.
   }
   {In this paper we report on multifrequency monitoring by the WEBT aimed at following the
   further evolution of the outburst in detail. In particular, we investigate the 
   expected correlation and time delays between the optical and radio emissions in order to derive 
   information on the variability mechanisms and jet structure.
   }
   {A comparison among the light curves at different frequencies is performed by means 
   of visual inspection and discrete correlation function, and the results are interpreted 
   with a simple model taking into account Doppler factor variations of geometric origin.
   }
   {The high-frequency radio light curves show a huge outburst starting during the dimming phase
   of the optical one and lasting more than 1 year. The first phase is characterized by a slow flux
   increase, while in early 2006 a major flare is observed. The lower-frequency radio light curves 
   show a progressively delayed and fainter event, which disappears below 8 GHz. 
   We suggest that the radio major peak is not physically connected with the 
   spring 2005 optical one, but it is actually correlated with a minor optical flare 
   observed in October--November 2005. This interpretation involves both an intrinsic and a geometric mechanism.
   The former is represented by disturbances travelling down the emitting jet, the latter being due to
   the curved-jet motion, with the consequent differential changes of viewing angles of the different emitting 
   regions.}
   {}

   \keywords{galaxies: active --
             galaxies: quasars: general --
             galaxies: quasars: individual: \object{3C 454.3} --
             galaxies: jets}

   \maketitle
%

\section{Introduction}

In May 2005 the quasar-type blazar \object{3C 454.3} (2251+158) was observed in an unprecedented luminous state 
from near-IR to hard X-ray frequencies. A big observing effort was spent to follow the outburst, 
involving both ground-based and space instruments \citep{fuh06,gio06,pia06,vil06}.
In particular, a large multiwavelength campaign was organized by the Whole Earth Blazar Telescope 
(WEBT)\footnote{{\tt http://www.to.astro.it/blazars/webt/}\\ see e.g.\ \citet{vil04b,vil04a,rai05,rai06b}.}, 
whose first results, up to September 2005, were reported and analysed by \citet{vil06}. 
A huge mm outburst followed the optical one, peaking in June--July 2005.
In the meantime the high-frequency (43--37 GHz) radio flux started to increase. 
VLBA observations at 43 GHz during the summer 
confirmed the brightening of the radio core and showed an increasing polarization.

Follow-up radio-to-optical observations by the WEBT continued until the end of the 
optical observing season, while a new campaign including three pointings by the
XMM-Newton satellite started just after and it is still ongoing.
In this paper we present part of these new observations with the aim of studying the correlation 
between the radio and optical emissions. 


\section{Observations and analysis}

\subsection{Optical and radio light curves}

   \begin{figure*}
   \centering
   \includegraphics[height=18cm]{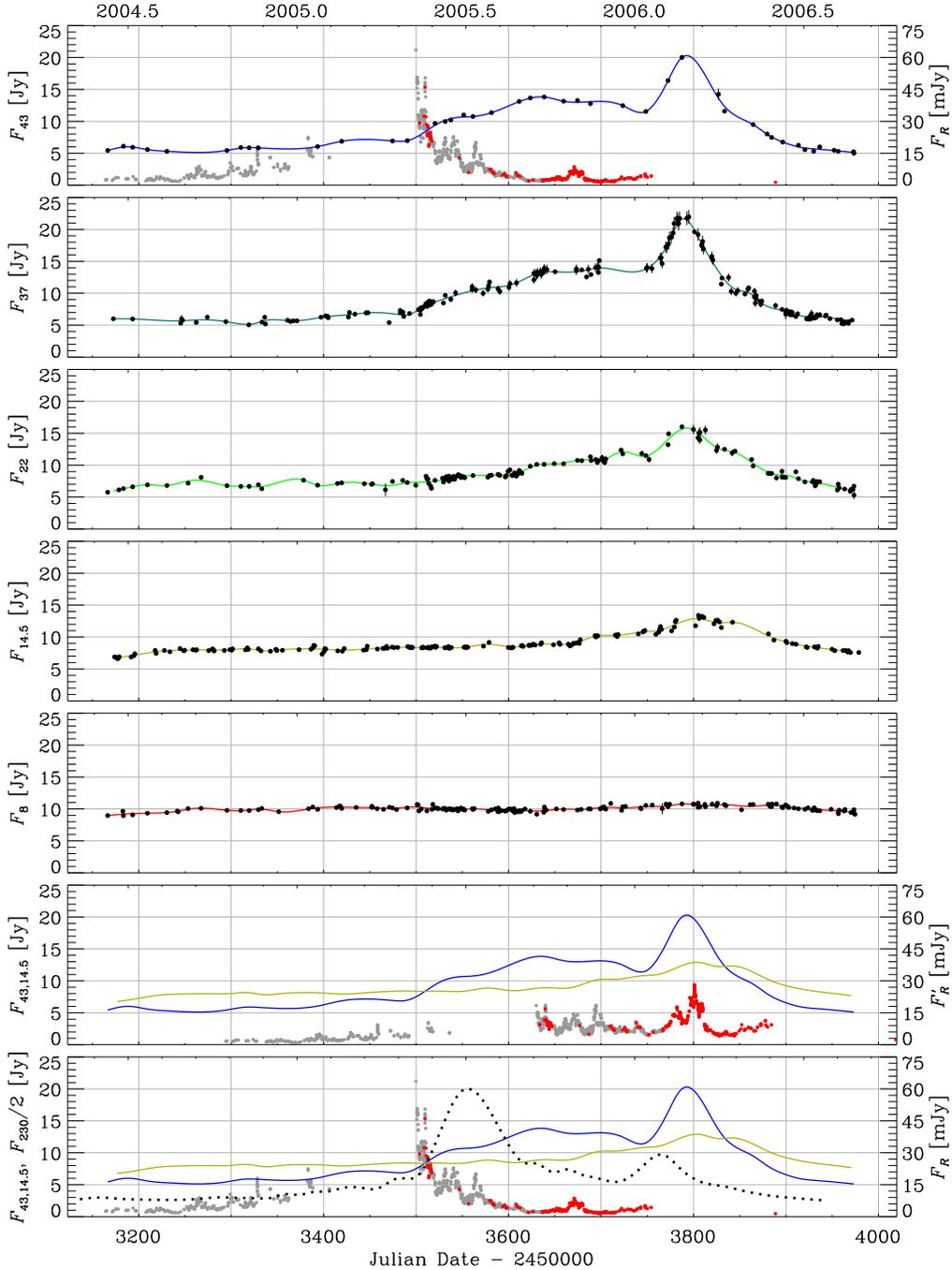}
   \caption{Optical and radio light curves of 3C 454.3 from June 2004 to the end of August 2006. 
   In the top and bottom panels the grey dots represent $R$-band flux densities from \citet{vil06},
   while red dots display new data. In the last but one panel the $R$-band light curve (grey and red dots)
   has been modified by ``rebeaming" and time shifting, as explained in the text. 
   The first five panels from top show the 43, 37, 22, 14.5, and 8 GHz light curves along with their
   15-day binned cubic spline interpolations. The 43 and 14.5 GHz splines are also reported
   in the last two panels for a comparison with the other light curves. In the bottom panel a sketch of
   the predicted 1 mm (230 GHz) light curve is also plotted as a dotted line.}
   \label{fig1}
   \end{figure*}

In the top panel of Fig.\ \ref{fig1} we show the $R$-band data taken by the WEBT 
from June 2004 to January 2006; grey dots refer to data already published in \citet{vil06}, while red 
dots represent the new data collected for this paper. The rightmost point at $\rm JD = 2453888.7$ (June 2, 2006)
indicates the low optical level at which the source was found after the 2006 solar conjunction.
These new optical data were taken at the 
Osaka Kyoiku,
Mt.\ Maidanak, 
Abastumani,
Crimean,
Torino,
Sabadell,
Roque de los Muchachos (KVA and NOT), 
and Mt.\ Lemmon Observatories.
In the same panel the 43 GHz radio data from June 2004 to the end of August 2006 are shown as black 
dots together with a cubic spline 
interpolation through the 15-day binned light curve (blue line). Analogously, the following four panels
display data and corresponding splines at 37, 22, 14.5, and 8 GHz. 
The points after $\rm JD = 2453644.5$ (end of September 2005) represent new data.
They were acquired at the Crimean (RT-22), Medicina,  Mets\"ahovi, Noto, SAO RAS (RATAN-600), and 
UMRAO Radio Observatories. Measurements from the VLA/VLBA Polarization Calibration Database\footnote{
{\tt http://www.vla.nrao.edu/astro/calib/polar/}} are also used.

In the $R$-band flux-density light curve one can recognize the rather slow rising phase
of the outburst started around $\rm JD = 2453250$, followed by a period of missing data because of the 2005
solar conjuction. After this, an unprecedentedly high flux density was observed, rapidly decreasing
towards the end of the outburst, which can be located around $\rm JD = 2453620$. The total outburst 
duration was thus about 1 year.

The 43 GHz light curve shows a double-humped outburst starting about 250 days
after the optical one, and representing the brightest event thus far recorded for this object
at high radio frequencies (see the historical light curves in \citealt{vil06}). 
The first hump ($\rm JD \sim 2453500$--2453750) is broader and lower, reaching at most $\sim 14$ Jy, 
while the second hump is more peaked, with a maximum of $\sim 20$ Jy. 
The whole event lasted a bit more than 400 days and, at first sight, it 
seems to mimic well the observed slow rise and fast drop of the optical outburst, 
with a lag of 250--300 days.

The 37 GHz flux density behaviour is similar, with the main difference that the first hump is 
substituted by a quasi-monotonic increase, which becomes flatter before the steep rise leading to
the maximum peak. This is even higher ($\sim 22$ Jy) than the 43 GHz one. The dates of start, peak, 
and end of the event seem to be practically identical to the previous ones.

In the other radio light curves, one can see similar features.
However, going towards lower frequencies, the variability amplitude becomes
smaller and smaller, while the starting point of the outburst, and to a much less extent also
the peak, are more and more delayed. In the 11 GHz light curve (not shown in the figure)
the outburst is still clearly observed, with a maximum flux density increase of more than 20\%,
while at 8 GHz the event is reduced to a barely visible $\sim 10\%$ flux density enhancement,
which begins about 200 days later than the highest-frequencies one.
The 5 GHz light curve (not shown in the figure) is almost completely flat, 
or even slightly decreasing, in the corresponding period.

\subsection{Cross-correlation between radio light curves}

In the following, we analyse the time lags mentioned above by means of the 
discrete correlation function (DCF; \citealt{ede88,huf92}).
We first investigate the delays among the starting points of the outburst at the
different radio wavelengths. To do this, we restrict the calculation to the period 
before the rise to the maximum peak, to avoid its strong imprint.

The DCF computed on this pre-maximum period ($\rm JD \la 2453750$) indicates
a delay of the 37 GHz starting point with respect to the 43 GHz one of 
5 days\footnote{Time lags are determined by taking the centroid of the 
DCF peaks \citep[see e.g.][]{pet01,rai03}. A precise estimate of the uncertainty for the
time lags is not easy, depending on the method used and parameters adopted. However, through Monte 
Carlo simulations, we found typical values around 5 days for almost all the time delays reported 
in this paper, but for the more uncertain lags of the outburst starting 
points at 22 and 14.5 GHz, having uncertainties of $\sim 10$--20 days.}. 
The DCF between the 37 and 22 GHz light curves shows two equivalent maxima; 
the first and broader one suggests a mean time lag of 25 days, 
while the second one indicates a 95 day delay.
Finally, the 14.5 GHz outburst would start 160 days later than the 37 GHz one.

\begin{figure}
  \resizebox{\hsize}{!}{\includegraphics{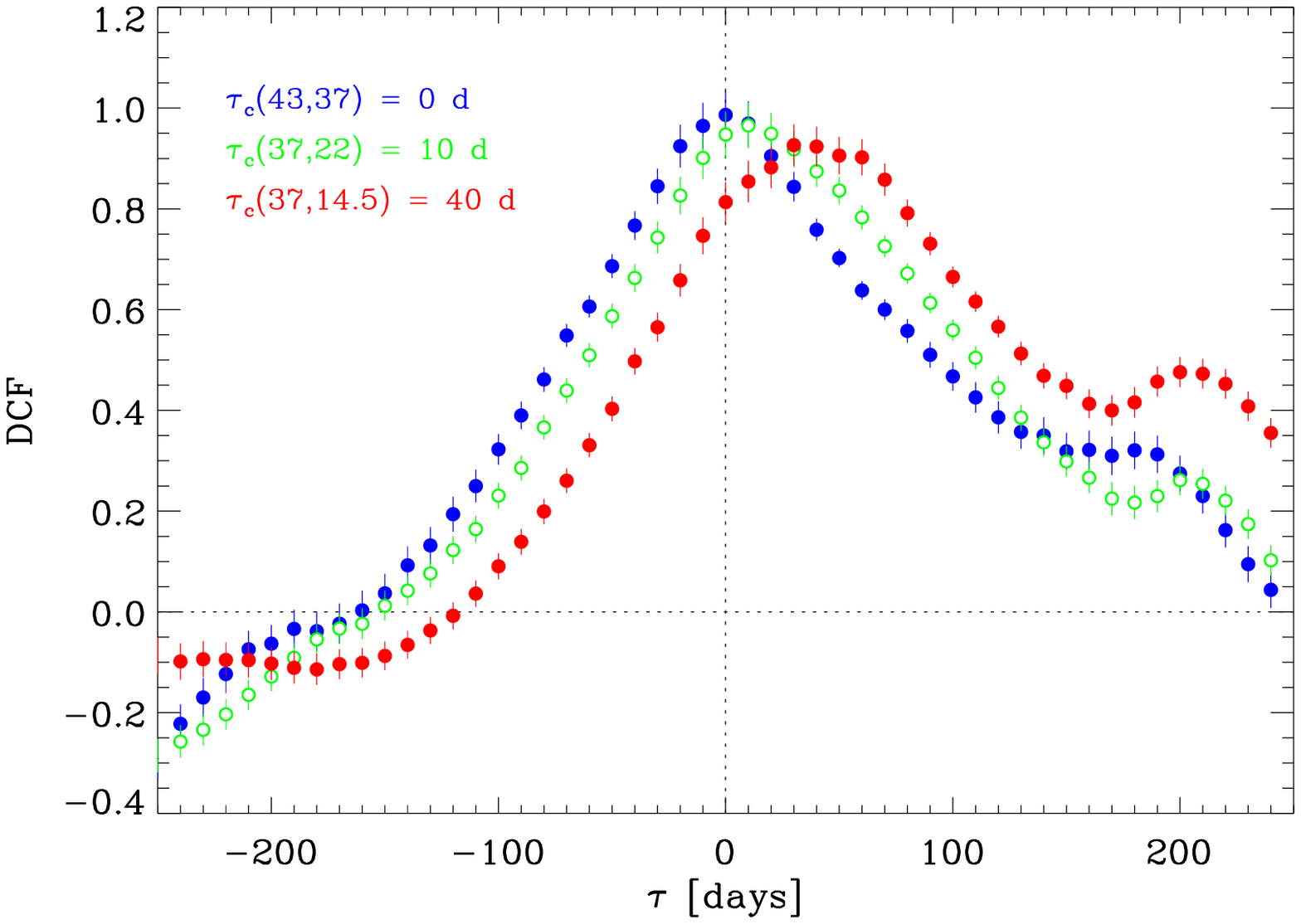}}
  \caption{Discrete correlation functions (DCFs) between radio light curves: 43 versus 37 GHz 
  (blue filled circles), 37 versus 22 GHz (green empty circles), and 37 versus 14.5 GHz
  (red filled circles).}
  \label{fig2}
  \end{figure}

Now we consider the whole 815 day period shown in Fig.\ \ref{fig1}; 
consequently, the DCF results will be dominated by the presence of the peaks.
The DCF applied to the spline interpolations leads to the same results as when using 
the original data points; hence, in Fig.\ \ref{fig2} we show the less noisy cross-correlations 
between splines. 
As one can see, the radio emissions from 43 to 14.5 GHz are all well correlated.
While no evident lag is present between the 43 and 37 GHz peaks (blue filled circles), 
the 22 GHz peak is delayed by 10 days (green empty circles), and 40 days separate the 14.5 GHz event
from the 37 GHz one (red filled circles).

The progressive delays of the outburst starting points can be interpreted in terms of a 
disturbance travelling down the jet through more and more transparent emitting regions.
In this view, the above results suggest that in the observer's frame
the jet becomes transparent quasi-simultaneously 
at 43 and 37 GHz. Then, some tens of days are needed to reach the region where 
it is barely transparent also to the 22 GHz emission, which however is maximally radiated 
about 100 days after the highest frequencies. 
A couple of months later also the 14.5 GHz emission can escape the jet.

However, in a simple model, one would expect that also the peaks and subsequent dimming phases
present comparable delays, while here they are found to be much shorter.
Actually, it seems that something occurred, making the outburst stop quasi-simultaneously 
at all radio frequencies.

\subsection{Cross-correlation between optical and radio light curves}

\begin{figure}
  \resizebox{\hsize}{!}{\includegraphics{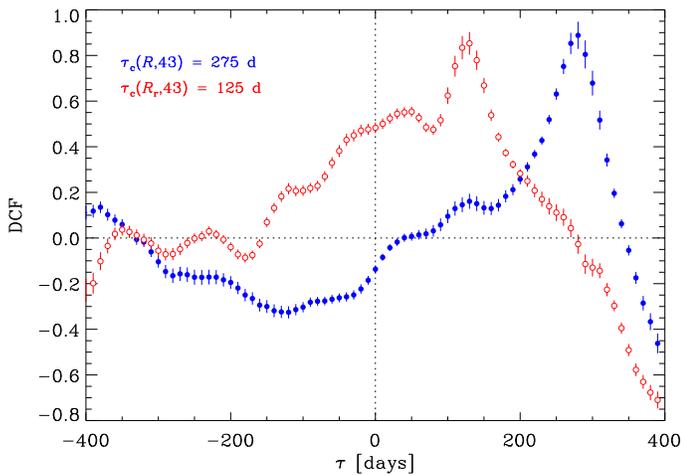}}
  \caption{Discrete correlation function (DCF) between the $R$-band and 43 GHz light curves 
  (blue filled circles) compared with the DCF between the ``rebeamed" $R$-band 
  flux densities and the 43 GHz light curve (red empty circles; see text for details).}
  \label{fig3}
  \end{figure}

As noticed in Sect.\ 2.1 from a visual inspection of Fig.\ \ref{fig1}, the optical and 43 GHz 
light curves seem to nicely correlate with a 250--300 day radio lag.
Indeed, the cross-correlation analysis yields a 275 day delay, as shown in Fig.\ \ref{fig3}, 
where the DCF between the 1-day binned $R$-band light curve
and the 43 GHz spline is plotted as blue filled circles\footnote{We chose the 43 GHz light curve instead 
of the better-sampled 37 GHz one because this latter misses an important feature like the first hump of 
the outburst mentioned in Sect.\ 2.1.}.

This long time delay is not expected, 
especially in view of the fact that the millimetric 
outburst peaked around mid 2005, i.e.\ shortly after the optical one \citep{vil06}.
Indeed, it is difficult to conceive why there should be
such a little time delay (2--3 months) between the optical and mm variations, 
and such a large lag (about 9 months) between the optical and radio ones.

In fact, \citet{vil06} claimed that the outburst had fully propagated to the high radio 
frequencies by the end 
of September 2005, corresponding to the first hump of the 43 GHz light curve and implying
an optical-radio delay of 4--5 months. Indeed, in the above optical-radio DCF 
(blue filled circles in Fig.\ \ref{fig3}) there is a little signal at about 130 days.
This signal remains also when the DCF is calculated on the data 
after $\rm JD = 2453600$ to exclude the optical outburst. Actually, it becomes the dominant one,
even if faint ($\rm DCF \sim 0.3$). It is due to the correlation between 
the minor optical flare peaking at $\rm JD \sim 2453670$ (see top panel of Fig.\ \ref{fig1})
with the maximum peak of the 43 GHz light curve.
In the following we thus consider the possibility that the 43 GHz maximum peak is not physically linked 
with the major optical outburst of spring 2005, but with the minor optical flare of October--November 2005.


\section{Discussion and conclusions}

\citet{vil06} suggested that the unprecedented luminosity of the spring 2005 optical
outburst was mainly due to a decrease of the viewing angle of the optically emitting jet region,
implying an increase of the beaming factor. In the same paper the authors argued also that
the jet should present some bending,
because of the lack of correlation between the historical
optical and radio light curves. Hence, 
the radio outburst which followed after 4--5 months 
(first hump of the 43 GHz light curve) most likely
occurred in a slightly misaligned outer region,
so that the radio flux was not so enhanced as the optical one.
At the time of the minor optical flare observed in October--November 2005
the jet must have turned, since the beaming of the optical radiation has decreased.
On the other hand, the radio emitting region seems to have reached the minimum viewing angle
at the time of the maximum peak.

This misalignment between the two regions makes the optical-radio
cross-correlation analysis rather difficult.
We thus investigate what happens when a correction for the misalignment is attempted,
by simulating an optical light curve suffering beaming conditions comparable with
the ones affecting the radio emission 130 days later, when the disturbances eventually reach
the high-frequency radio emitting region (see Sect.\ 2.3).

The ``true" optical flux densities $F_R$ are transformed into ``rebeamed", simulated ones, $F'_R$,
under the following assumptions: 
i) the observed flux density is proportional to $\delta^3$, 
where $\delta=[\gamma(1-\beta \cos\theta)]^{-1}$ is the Doppler (beaming) factor
and $\gamma$ is the Lorentz factor\footnote{We neglect here the minor Doppler effects 
on time intervals and frequencies.};
ii) the minimum viewing angle $\theta_{\rm min}$ of the optically emitting region 
is achieved at $t=t_0$ for the ``true" optical light curve, and 
at $t=t'_0$ in the case of the simulated one;
iii) the viewing angle increases both forward and backward in time with the square
root of the time elapsed since $t_0$,
$\theta(t)=\theta_{\rm min} + \sqrt{|t-t_0| / t_{\rm s}}$, and similarly for the 
``rebeamed" case.

In the last but one panel of Fig.\ \ref{fig1} we show the ``rebeamed" optical light curve
(grey and red dots) obtained with the model parameters set to:
$\gamma=10$,
$\theta_{\rm min}=5 \degr $, 
$t_0=2453490$, 
$t'_0=2453650$, 
$t_{\rm s}=14$ days\footnote{While the resulting light curve is not very sensitive to changes in $\gamma$,
$\theta_{\rm min}$, and $t_{\rm s}$, it is obviously strongly affected by the values of $t_0$ and $t'_0$, 
which have been chosen close to the major (spring) and minor (autumn) optical events, respectively.}.
Moreover, the simulated light curve has been shifted in time by 130 days in order to make
the comparison with the 43 GHz spline easier.
One can see that the general trend of the two light curves is quite similar, with
the main difference that the 43 GHz variations appear to be smoother, as expected, 
since the radio emission is thought to come from a larger region.
The result of their cross-correlation is shown in Fig.\ \ref{fig3} as red empty circles,
indicating a strong correlation with a radio delay of 125 days.

In other words, the radio delays come from the concomitance of two mechanisms acting on
different time scales.
The first mechanism is the perturbation travelling along the inhomogeneous jet,
which causes the $\sim 125$ day radio lag. 
The other one is the ``turning" of the curved jet, producing a differential variation of the viewing
angle of the different emitting regions. This mechanism is responsible for the 275 day separation
between the times of the optical and radio maximum beaming.
In this framework the difference between the long delays
of the radio outburst starting points and the shorter delays of the peaks discussed in Sect.\ 2.2
can be understood in terms of the different orientation of the radio emitting region at 
the times of the passage of the two perturbations, implying a greater contraction of the time scales
when the viewing angle is smaller.

After the maximum peak, the radio flux drops rapidly at all the higher frequencies, and
no flux enhancement is seen at the lower frequencies, as one would expect from the transit
of the disturbance through outer regions. 
The reason may be that these jet regions are bent in such a way that the radiation emitted there 
is beamed elsewhere.
Moreover, from the light curves in Fig.\ \ref{fig1} one can see that the high-frequency
radio spectrum is strongly inverted during the outbursts,
suggesting that these events occurred inside the radio core, where low radio 
frequencies are still strongly absorbed.
In other words, after mid 2006 the disturbances would be travelling in the misaligned jet region, 
i.e.\ out of the core (see the discussion by \citealt{bac06} on the VLBA maps of BL Lacertae). 
Eventually they may become again visible when passing through outer regions 
having again a small viewing angle, and consequently appear as new radio components. 
Indeed, VLBA radio maps at 43 GHz taken in April and August 2006 do not show 
any new component yet, even if a slight elongation of the core
in the jet direction is visible in the August map, i.e.\ towards the end of our observing
period (Marscher et al., in preparation).

Finally, the above model can lead to a prediction for the behaviour of the millimetric emission
in the same period.
A reasonable hypothesis is that at these wavelengths the outburst development
is intermediate between those observed in the optical and radio bands, with a first very strong mm 
outburst (as already reported by \citealt{vil06}), followed by a second, less prominent one.
In the last panel of Fig.\ \ref{fig1} the dotted line represents a sketch of the predicted
1 mm (230 GHz) light curve. It has been obtained as the average between
the 43 GHz spline and a curve representing the optical data, shifted by 125 days.
This latter curve is a cubic spline interpolation through the 15-day binned $R$-band
flux densities, scaled as shown in the same panel to make them comparable with the radio ones.
Then, the so obtained 1 mm curve has been normalized to have a maximum peak of about 40 Jy, 
and it has been shifted by 30 days back in time to make the peak occur in mid 2005, 
to match the observations quoted by \citet{vil06}.

In summary, we are envisaging a scenario where the optically emitting region of a curved jet 
acquires its minimum viewing angle in spring 2005 ($\rm JD \sim 2453490$), 
so that some disturbance(s) travelling in that
region produces a maximally Doppler-enhanced outburst. 
Soon after also the mm region is affected by similar
intrinsic (perturbation) and geometric (minimum viewing angle) 
conditions, even if this latter condition is probably reached with some delay.
In the meanwhile some other perturbations 
(minor flares up to $\rm JD \sim 2453575$ in the optical light curve)
are crossing the optical region, which is becoming
less and less well aligned with the line of sight. 
Most likely, some of these disturbances enter the mm region when it is still well oriented, 
thus producing a mm outburst more extended in time (probably more extended than in our toy-model
prediction). 
The first disturbance is now crossing the 43--37 GHz region, followed by the other ones. This train 
gives rise to the broad and modulated hump lasting about 250 days ($\rm JD \sim 2453500$--2453750).
The lower-frequency emitting regions are also progressively perturbed. 
A new perturbation crosses the now misaligned optical region (flare around $\rm JD = 2453670$), 
and produces a mild-intensity outburst when reaching the less misaligned mm region.
Finally, it enters the high-frequency radio regions when their viewing angle is minimum, and it is 
Doppler-enhanced into an exceptionally bright event. All this happens in the VLBI radio core, where low 
frequencies are still well absorbed, and the outburst radio spectrum is strongly inverted. Only at the 
end of the outburst it gets softer, when the last disturbance leaves the highest-frequency radio regions.

Optical-to-radio monitoring by the WEBT is continuing to follow the source post-outburst phases.

\begin{acknowledgements}
We thank Alan Marscher for sharing useful information concerning changes in the VLBA images of 3C 454.3, 
and Philip Hughes for useful discussion.
This work is partly based on observations made with the Nordic Optical Telescope, operated
on the island of La Palma jointly by Denmark, Finland, Iceland,
Norway, and Sweden, in the Spanish Observatorio del Roque de los
Muchachos of the Instituto de Astrof\'{\i}sica de Canarias.
It is partly based also on observations with the Medicina and Noto
telescopes operated by INAF - Istituto di Radioastronomia.
We thank the staff at the Medicina and Noto radio observatories for
their help and support during the observations.
This research has made use of data from the University of Michigan Radio Astronomy Observatory,
which is supported by the National Science Foundation and by funds from the University of Michigan.
This work was partly supported by the Italian Space
Agency (ASI) under contract ASI/INAF I/023/05/0. The St.\ Petersburg team  
acknowledges support from Russian Federal Program for Basic Research under grant 05-02-17562.
RATAN-600 observations were partly supported by the
Russian Foundation for Basic Research grant 05-02-17377.
This project was done while YYK was a Jansky fellow of the National
Radio Astronomy Observatory and a research fellow of the Alexander von
Humboldt Foundation.
\end{acknowledgements}

\end{document}